\documentclass[twocolumn,aps,showpacs,superscriptaddress,tightenlines,a4paper]{revtex4}

\usepackage{amssymb}
\usepackage{amsmath}
\usepackage{bbold}
\usepackage{mathrsfs}
\usepackage{paralist}
\usepackage{graphicx}
\usepackage{xcolor}

\begin{document}
%
% TITLE
%
	\title{The cutoff-dependence of the Casimir force within an inhomogeneous medium}
%
% AUTHORS
%

	\author{S. A. R. Horsley}
	\affiliation{Electromagnetic and Acoustic Materials Group, Department of Physics and Astronomy, University of Exeter, Stocker Road, Exeter, EX4 4QL, UK}
	\email{s.horsley@exeter.ac.uk}
	\author{W. M. R. Simpson}
	\affiliation{School of Physics and Astronomy, University of St Andrews, North Haugh, St Andrews, KY16 9SS, UK}
%	\email{wmrs2@st-andrews.ac.uk}
	\affiliation{Faculty of Physics, Weizmann Institute of Science, 234 Herzl St., Rehovot 76100, Israel}

%
% ABSTRACT
%
	\begin{abstract}
		We consider the ground state energy of the electromagnetic field in a piston geometry. In the idealised case, where the piston and the walls of the chamber are taken as perfect mirrors, the Casimir pressure on the piston is finite and independent of the small scale physics of the media that compose the mirrors; the Casimir-energy of the system can be regularised and is cutoff-independent. Yet we find that, when the body of the piston is filled with an inhomogeneous dielectric medium, the Casimir energy is cutoff-dependent, and the value of the pressure is thus inextricably dependent on the detailed behaviour of the mirror and the medium at large wave-vectors. This result is inconsistent with recent proposals for regularising Casimir forces in inhomogeneous media. 
	\end{abstract}
	%
	% Quantum optics (42.50.-p) Quantum noise (42.50.Lc)
	%
	\maketitle
	\bibliographystyle{unsrt}
%
%: Introduction
%	

\section{Introduction}
	\par
	It is now a well-known and experimentally supported fact that two parallel uncharged mirrors at zero temperature will exert an attractive force upon each other, as Casimir predicted~\cite{Casimir1948}---a force arising from the ground state properties of the electromagnetic field. The theory has been made quantitative, applying to media described by \(\epsilon(\omega)\) and \(\mu(\omega)\) satisfying the Kramers--Kr\"{o}nig relations~\cite{lifshitz1955,DzyLifPit1961,volume9,philbin2011}, and Casimir forces have been calculated for a variety of systems and geometries (e.g. see~\cite{bordag2009,dalvit2011}). The purpose of the following discussion is to demonstrate that there remains a problem within the theory of the Casimir effect for the case of inhomogeneous media, in spite of recent efforts to solve or circumvent it~\cite{rosa2011,goto2012}.  In a separate paper~\cite{simpson2013} we have shown that this problem is inherent within the Lifshitz theory of the Casimir effect, where dispersion and dissipation are properly included; the Casimir-stress in such cases is infinite and resists regularisation.
	\par
But this problem is not peculiar to Lifshitz theory. Here we show that, even when we make the most basic attempt to mimic dielectric media with non--dispersive boundary conditions---considering only a simple energy summation of the field modes---the Casimir force is infinite in the limit where the regularization (cut--off) tends to infinite frequency.  It seems apparent that we cannot obtain an expression for the zero point force that is independent of the choice of cut off in the energy summation.
	
%
%: Section 1 - The Casimir piston
%	
\section{The Casimir piston\label{section-1}}
\par
\begin{figure}[h!]
	\includegraphics[width=6.5cm]{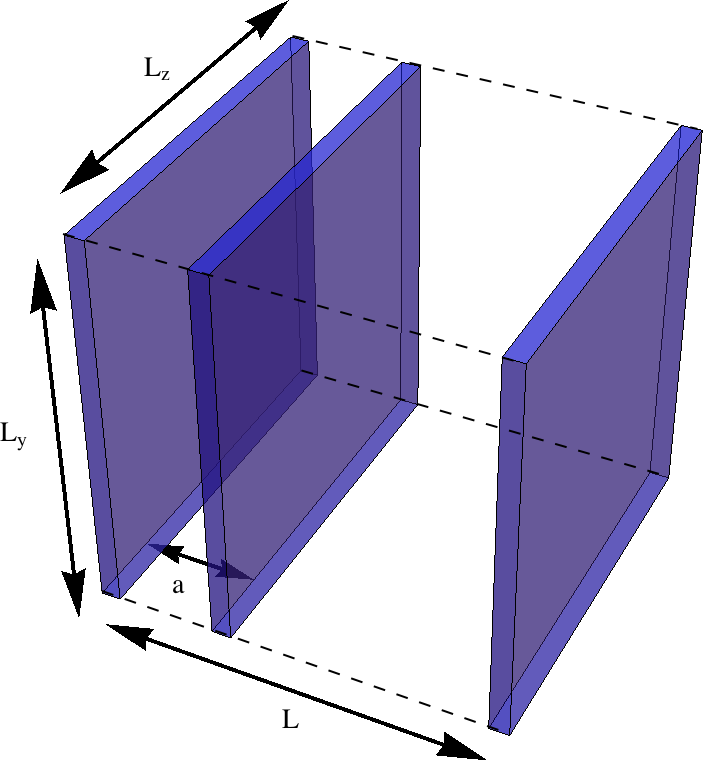}
	\caption{Schematic of the Casimir piston.  Two fixed mirrors are positioned at \(x=0\) and \(x=L\), enclosed by reflecting walls at \(y=\pm L_{y}/2\) and \(z=\pm L_{z}/2\) (dashed lines).  Within the chamber is vacuum, and a moveable mirror at \(x=a\).  While the ground state energy of the electromagnetic field in this system depends on the small scale physics of the mirrors, the part of it dependending on \(a\) does not.\label{figure-1}}	
\end{figure}

	We first briefly illustrate the principles of our main calculation in an idealized system where we know there exists a finite expression for the Casimir force, independent of the microscopic physics of the bodies involved.   The chosen system is a cavity of length \(L\), and cross-sectional area \(A=L_{y}L_{z}\), divided by a mirror at a distance \(a\) from the left--most cavity wall (figure~\ref{figure-1}).  Even at \(T=0\,\text{K}\) the moveable mirror situated at \(x=a\) is subject to a force due to the dependence of the field energy on \(a\).
\par
	We can formally write the ground state energy of the system as the sum of all possible contributions of \(\hbar\omega/2\),
\begin{equation}
	E=\frac{\hbar}{2}\sum_{m,p,q,\lambda}\left[\omega_{m,p,q,\lambda}^{L}+\omega_{m,p,q,\lambda}^{R}\right],\label{ground-state-energy}
\end{equation}
where \(\lambda\in\{1,2\}\) indicates the polarization, and the range of summation for each of the indices runs over the allowed modes.  The eigenfrequencies of the cavities to the left (\(L\)) and right (\(R\)) of \(a\) are independent of the polarization when \(m>0\),
\begin{align}
	\omega_{m,p,q}^{L}&=\pi c\sqrt{\frac{m^{2}}{a^{2}}+\frac{p^{2}}{L_{y}^{2}}+\frac{q^{2}}{L_{z}^{2}}}\nonumber\\
	\omega_{m,p,q}^{R}&=\pi c\sqrt{\frac{m^{2}}{(L-a)^{2}}+\frac{p^{2}}{L_{y}^{2}}+\frac{q^{2}}{L_{z}^{2}}},\label{eigenfrequencies}
\end{align}
	and when \(m=0\), the \(\lambda=1\) polarization is not an allowed mode of the cavity.  Given the behaviour of (\ref{eigenfrequencies}), (\ref{ground-state-energy}) is not a meaningful expression: the summand becomes ever larger as \(m\), \(p\) \& \(q\) are increased, and the sum diverges.  This divergence is due to the artificial assumption that there exist mirrors that act at all frequencies.  To fix this, a factor is inserted into the eigenfrequencies (\ref{eigenfrequencies}) to make the sum converge,
\begin{equation}
	\tilde{E}=\hbar\sum_{m,p,q}\left[\omega_{m,p,q}^{L}e^{-\xi\omega_{m,p,q}^{L}/c}+\omega_{m,p,q}^{R}e^{-\xi\omega_{m,p,q}^{R}/c}\right],\label{regularized-energy}
\end{equation}
where \(\xi\) is a free parameter, and a factor of two comes from the symmetry of the system with respect to polarization.  This modified expression for the energy can evidently no longer be considered as the total energy of the system of field plus mirrors.  Taken literally as such, (\ref{regularized-energy}) would imply that the eigenfrequencies of the field eventually all tend to zero, an assumption for which there is no obvious motivation.  Instead, (\ref{regularized-energy}) should be interpreted as the part of the total energy (\ref{ground-state-energy}) associated with the configuration of mirrors in the cavity (the energy available to do work on the mirrors).  The exponential factors then amount to a model for the dispersive behaviour of the mirrors: i.e. at sufficiently high frequencies/wave--vectors the mirrors become transparent, and after this the total energy does not depend on their configuration.  Note that the factor of two in (\ref{regularized-energy}) takes into account the sum over polarization.  This is not correct for the \(m=0\) mode.  However, the energy of this mode is independent of \(a\).
%
%: Figure 1 - The Casimir piston
%
\par
	To explicitly evaluate (\ref{regularized-energy}) we take the limit \(L_{y}/a,L_{z}/a\to\infty\), where the summation over \(m\) and \(p\) can be converted into an integration over \(k_{y}\) and \(k_{z}\), \(\Delta k_{i}=\pi/L_{i}\to dk_{i}\) (\(i\) is either \(y\) or \(z\)).  After an integration over the angle, \(\theta\) of \(\boldsymbol{k}_{\parallel}=k_{y}\hat{\boldsymbol{y}}+k_{z}\hat{\boldsymbol{z}}=k_{\parallel}[\cos(\theta)\hat{\boldsymbol{y}}+\sin(\theta)\hat{\boldsymbol{z}}]\), the energy per unit area of the piston configuration is found to be
\begin{multline}
	\frac{\tilde{E}}{A}=\frac{\hbar c}{2\pi}\sum_{m=0}^{\infty}\int_{0}^{\infty}k_{\parallel}dk_{\parallel}\sqrt{\left(\frac{m\pi}{a}\right)^{2}+k_{\parallel}^{2}}\,e^{-\xi\sqrt{\left(\frac{m\pi}{a}\right)^{2}+k_{\parallel}^{2}}}\\
	+a\to L-a,\label{energy-per-area}
\end{multline}
	where \(a\to L-a\) indicates a repetition of the previous expression with `\(a\)' replaced everywhere by `\(L-a\)'.  To (\ref{energy-per-area}) we apply the identity,
\begin{multline}
	\frac{d}{d k_{\parallel}}\left[\left(\xi^{-1}\sqrt{\left(\frac{m\pi}{a}\right)^{2}+k_{\parallel}^{2}}+\xi^{-2}\right)e^{-\xi\sqrt{\left(\frac{m\pi}{a}\right)^{2}+k_{\parallel}^{2}}}\right]\\
	=-k_{\parallel}e^{-\xi\sqrt{\left(\frac{m\pi}{a}\right)^{2}+k_{\parallel}^{2}}},
\end{multline}
	yielding,
	\begin{equation}
		\frac{\tilde{E}}{A}=\frac{\hbar c}{\pi}\left[\frac{1}{\xi^{3}}-\frac{1}{\xi^{2}}\frac{d}{d\xi}+\frac{1}{2\xi}\frac{d^{2}}{d\xi^{2}}\right]\sum_{m=0}^{\infty}e^{-\xi m\pi/a}+a\to L-a.\label{geometric-series}
	\end{equation}
	The summation within (\ref{geometric-series}) is evidently a geometric series, which can be evaluated, \(\sum_{m=0}^{\infty}e^{-\alpha m}=1/(1-e^{-\alpha})\), giving,
	\begin{multline}
		\frac{\tilde{E}}{A}=\frac{\hbar c}{2\pi}\left[\frac{1}{\xi^{3}}-\frac{1}{\xi^{2}}\frac{d}{d\xi}+\frac{1}{2\xi}\frac{d^{2}}{d\xi^{2}}\right]e^{\xi\pi/2a}\text{cosech}(\xi\pi/2a)\\
		+a\to L-a.\label{energy-per-A}
	\end{multline}
	The introduction of the exponential factor into (\ref{regularized-energy}) represents an extremely artificial model for the behaviour of the mirrors at high frequencies.  We therefore separate the energy into those parts that depend on \(\xi\), and those that do not.  In anticipation of taking the limit \(\xi\to0\), the quantity to the right of the square brackets in (\ref{energy-per-A}) is expanded as far as \(\xi^{3}\),
	\begin{equation}
		e^{\xi\pi/2a}\text{cosech}(\xi\pi/2a)\sim\frac{2a}{\xi\pi}+1+\frac{1}{3}\left(\frac{\xi\pi}{2a}\right)-\frac{1}{45}\left(\frac{\xi\pi}{2a}\right)^{3}.\label{expansion-1}
	\end{equation}
	Inserting (\ref{expansion-1}) into (\ref{energy-per-A}) gives finally,
	\begin{equation}
		\frac{\tilde{E}}{A}=\hbar c\left[\frac{3L}{\pi^{2}\xi^{4}}+\frac{1}{\pi\xi^{3}}-\frac{\pi^{2}}{720a^{3}}-\frac{\pi^{2}}{720(L-a)^{3}}\right].\label{casimir-energy}
	\end{equation}
	As expected, the energy becomes increasingly large as \(\xi\to0\).  However, the part of the energy depending on the position of the mirror is independent of \(\xi\).  One may interpret this to mean that so long as \(\xi/a\) is negligibly small (i.e. neglecting the positive powers of \(\xi\) in (\ref{casimir-energy}) is legitimate), the part of the energy that depends on \(a\) is independent of how the mirror becomes transparent at high frequencies/wave--vectors.  Taking the derivative of (\ref{casimir-energy}) with respect to \(a\) then yields the usual finite and \(\xi\) independent pressure,
	\[
		\frac{F}{A}=-\frac{1}{A}\frac{\partial\tilde{E}}{\partial a}=\frac{\hbar\pi^{2} c}{240 (L-a)^{4}}-\frac{\hbar\pi^{2} c}{240 a^{4}},
	\]
	so that the Casimir force on the mirror pulls it towards the closer of the two end walls of the chamber.
%
%:  Section 2 - The inhomogeneous Casimir piston
%	
\section{The Inhomogeneous Casimir Piston\label{section-2}}
\par
	The above procedure for making (\ref{ground-state-energy}) convergent yields terms dependent on \(a\) that are all finite or zero in the limit \(\xi\to0\).  Yet it is not obvious that this fortuitous situation occurs for fundamental reasons.  Suppose we take the same cavity with an inhomogeneous medium within the chamber: are the divergent terms in \(\tilde{E}\) still independent of \(a\)?
\par
\begin{figure}[h!]
	\includegraphics[width=8cm]{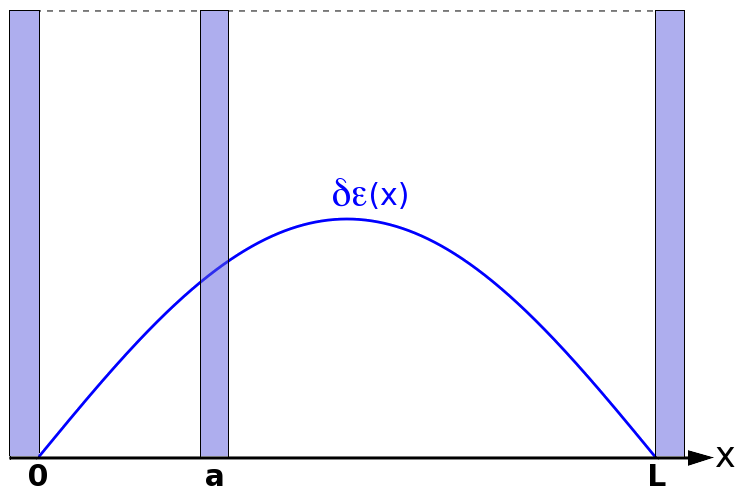}
	\caption{As in figure~\ref{figure-1}, we have a perfectly reflecting rectangular chamber of length \(L\) and cross sectional area, \(A\).  Within the chamber is a mirror at \(x=a\), surrounded by an inhomogeneous dielectric with \(\epsilon\) and \(\mu\) given by (\ref{epsilon-mu}).  We look for the dependence of the energy of this system on \(a\).\label{figure-2}}
\end{figure}

	Within the chamber we assume a permeability and permittivity given by (e.g. see figure~\ref{figure-2}),
\begin{align}
	\mu(x)&=\mu_{0}\nonumber\\
	\epsilon(x)&=\epsilon_{0}\left[1+\delta \epsilon(x)\right].\label{epsilon-mu}
\end{align}
	It is imagined that the medium with permittivity and permeability given by (\ref{epsilon-mu}) is a rigid body.  Is it possible to express the force holding the mirror fixed at \(a\) (opposing the Casimir force) in a form that depends only on \(\epsilon\) and \(\mu\)?
%
%: Figure 2 - the inhomogeneous Casimir piston
%
\par
	As in the previous section, and in common with Casimir's original calculation~\cite{Casimir1948}, we make the artificial assumption that \(\epsilon\) and \(\mu\) are independent of frequency.  This needs some justification.  These effects are of course fundamental to the interaction of light and matter~\cite{volume8}.  To properly account for them requires Lifshitz theory~\cite{lifshitz1955,DzyLifPit1961,volume9,philbin2011}, and in separate work we have shown that finite answers cannot be obtained from this theory in the above case~\cite{simpson2013}.  The purpose of this calculation is to examine whether this problem is peculiar to Lifshitz theory, or still present in a much more naive theory where the propagation speed of light is simply varied from point to point, and not as a function of frequency.
\par
	The situation is slightly complicated in comparison to section~\ref{section-1}, due to the fact that the two polarizations do not behave degenerately in the medium defined by (\ref{epsilon-mu}).  Within the cavity, the electromagnetic field obeys,
\begin{equation}
	\boldsymbol{\nabla}\boldsymbol{\times}\boldsymbol{\nabla}\boldsymbol{\times}\boldsymbol{E}_{m,\lambda}-\frac{\omega_{m,\lambda}^{2}}{c^{2}}[1+\delta\epsilon(x)]\boldsymbol{E}_{m,\lambda}=0,\label{wave-eqn}
\end{equation}
where \(m\) labels the spatial dependence of the mode, and \(\lambda\in\{1,2\}\) the polarization.  When \(\delta\epsilon=0\), the modes in the region \(x\in[0,a]\) are given by,
\begin{equation}
	\boldsymbol{E}_{m,1}^{(0)}=\hat{\boldsymbol{x}}\boldsymbol{\times}\hat{\boldsymbol{k}}_{\parallel}\sqrt{\frac{2}{aA}}\sin{(m\pi x/a)}e^{i\boldsymbol{k}_{\parallel}\boldsymbol{\cdot}\boldsymbol{x}},\label{mode-1}
\end{equation}
and,
\begin{multline}
	\boldsymbol{E}_{m,2}^{(0)}=\sqrt{\frac{2/aA}{k_{\parallel}^{2}+(m\pi/a)^{2}}}\bigg[k_{\parallel}\hat{\boldsymbol{x}}\cos(m\pi x/a)\\
	-i\hat{\boldsymbol{k}}_{\parallel}\left(\frac{m\pi}{a}\right)\sin(m\pi x/a)\bigg]e^{i\boldsymbol{k}_{\parallel}\boldsymbol{\cdot}\boldsymbol{x}}\label{mode-2}.
\end{multline}
where we assume the limit of \(L_{y}/a,L_{z}/a\to\infty\) as in section~\ref{section-1}.  After substituting, \(a\to L-a\), the modes in the region \(x\in[a,L]\) are also given by (\ref{mode-1}--\ref{mode-2}).  All modes are normalised over the volume, \(V\), of each region of the chamber, \(\int_{V}|\boldsymbol{E}_{m,\lambda}^{(0)}|^{2}d^{3}\boldsymbol{x}=1\).
\par
	We consider the case where \(\delta\epsilon(x)\ll1\), and write the frequency of each eigenmode as \(\omega_{m,\lambda}=\omega_{m,\lambda}^{(0)}+\omega_{n,\lambda}^{(1)}\), and the field as \(\boldsymbol{E}_{m,\lambda}=\boldsymbol{E}_{m,\lambda}^{(0)}+\boldsymbol{E}_{m,\lambda}^{(1)}\), where the quantities containing a superscript `\((1)\)' are supposed linear in \(\delta\epsilon\).  To first order in \(\delta\epsilon\), (\ref{wave-eqn}) is
\begin{multline}
	\boldsymbol{\nabla}\boldsymbol{\times}\boldsymbol{\nabla}\boldsymbol{\times}\boldsymbol{E}_{m,\lambda}^{(1)}-\frac{1}{c^{2}}\bigg[2\omega_{m,\lambda}^{(1)}\omega_{m,\lambda}^{(0)}\boldsymbol{E}_{m,\lambda}^{(0)}\\
	+\omega_{m,\lambda}^{(0)2}\boldsymbol{E}_{m,\lambda}^{(1)}+\omega_{m,\lambda}^{(0)2}\boldsymbol{E}_{m,\lambda}^{(0)}\delta\epsilon(x)\bigg]=0.\label{first-order-wave-eqn}
\end{multline}
Multiplying (\ref{first-order-wave-eqn}) on the left by \(\boldsymbol{E}_{m,\lambda}^{(0)\star}\) and integrating over \(V\) (which could either be the left or right region of figure~\ref{figure-2},) after an integration by parts we find,
\begin{equation}
	\omega_{m,\lambda}^{(1)}=-\frac{1}{2}\omega_{m,\lambda}^{(0)}\int\left|\boldsymbol{E}_{m,\lambda}^{(0)}\right|^{2}\delta\epsilon(x)d^{3}\boldsymbol{x},\label{first-order-correction}
\end{equation}
which is the standard expression for the first order perturbation of the eigenfrequencies of an optical cavity (see e.g.~\cite{pozar1998,johnson2002}).  For the particular case,
 \begin{equation}
 	\delta\epsilon(x)=\alpha\sin(\pi x/L)\label{perturbation-exp},
\end{equation}
we find after inserting (\ref{mode-1}--\ref{mode-2}) into (\ref{first-order-correction}), that in the left most portion of the piston with \(\lambda=1\),
	\begin{equation}
		\omega_{m,1}^{L\,(1)}=\omega_{m,1}^{L\,(0)}\frac{\alpha L\left[1-\cos(\pi a/L)\right]}{2\pi a}\frac{[2mL/a]^2}{1-[2mL/a]^2},\label{left-perturbation-1}
	\end{equation}
	and \(\lambda=2\),
	\begin{multline}
		\omega_{m,2}^{L\,(1)}=-\omega_{m,2}^{L\,(0)}\frac{\alpha L\left[1-\cos(\pi a/L)\right]}{2\pi a}\\
		\times\bigg\{1-\frac{\left(\frac{m\pi}{a}\right)^{2}-k_{\parallel}^{2}}{\left(\frac{m\pi}{a}\right)^{2}+k_{\parallel}^{2}}\frac{1}{1-[2mL/a]^{2}}\bigg\}\label{left-perturbation-2},
	\end{multline}	
	after substituting \(a\to L-a\) into (\ref{left-perturbation-1}--\ref{left-perturbation-2}), one has the expressions for the right hand side of the cavity.  The above shifts in the eigenfrequencies are independent of the angle, \(\theta\) of \(\boldsymbol{k}_{\parallel}\).  We can therefore perform the integration over \(\theta\) in the mode summation.  The change in energy per unit area due to the presence of the medium within the piston is computed, and takes the form,
	\begin{multline}
		\frac{\Delta\tilde{E}}{A}=\frac{\hbar}{4\pi}\sum_{\lambda,m}\int_{0}^{\infty}k_{\parallel}d k_{\parallel}\bigg[\omega_{m,\lambda}^{L\,(1)}e^{-\xi\omega_{m,\lambda}^{L\,(0)}/c}\\
		+\omega_{m,\lambda}^{R\,(1)}e^{-\xi\omega_{m,\lambda}^{R\,(0)}/c}\bigg],\label{energy-change}
	\end{multline}
	where we use the same regularizing function as in the unperturbed case.  The integration over \(k_{\parallel}\) is performed using the same techniques as in section~\ref{section-1}.  The results are given by (\ref{integral-1}--\ref{integral-2}) in appendix~\ref{appendix-A}.  In addition to terms proportional to geometric series' (c.f. (\ref{geometric-series})), the resulting sum over \(m\) contains terms of the form,
	\begin{equation}
		\Phi(e^{-\xi\pi/a},1,\upsilon)=\sum_{m=0}^{\infty}\frac{e^{-\xi m\pi/a}}{m+\upsilon}.\label{lerch-function}
	\end{equation}
	The quantity \(\Phi\) is known as the Lerch function~\cite{bateman1953,gradshteyn2007}.  Applying the notation of (\ref{lerch-function}), we can perform the summation, with the result,
\begin{widetext}
	\begin{multline}
		\frac{1}{A}\frac{d\tilde{E}}{d\alpha}=\frac{\hbar cL\left[1-\cos(\pi a/L)\right]}{4\pi^{2}a}\bigg\{\left(\frac{1}{\xi^{2}}\frac{d}{d\xi}-\frac{1}{\xi^{3}}-\frac{1}{2\xi}\frac{d^{2}}{d\xi^{2}}\right)e^{\xi\pi/2a}\text{cosech}(\xi\pi/2a)\\
		+\frac{a}{4L\xi}\frac{d^{2}}{d\xi^{2}}\left[\Phi(e^{-\xi\pi/a},1,a/2L)-\Phi(e^{-\xi\pi/a},1,-a/2L)\right]\bigg\}+a\to L-a,\label{energy-derivative}
	\end{multline}
\end{widetext}
	where we have converted the approximate expression, (\ref{energy-change}) into an exact relation through taking the limit \(\alpha\to d\alpha\) (as is done e.g. in~\cite{johnson2002}).  Equation (\ref{energy-derivative}) represents the rate of change of \(\tilde{E}\) as the amplitude of (\ref{perturbation-exp}) is increased from zero to \(d\alpha\).  Again we consider the limit \(\xi\to0\), and make use of the following series expansion of the Lerch function~\footnote{An examination of~\cite{gradshteyn2007} makes it appear as though the expansion (\ref{lerch-expansion}) does not apply to the case when the middle index equals unity.  However~\cite{bateman1953} shows that it does hold in this case, it is only that the expansion can be simplified to a hypergeometric function (we do not require this simplification).  We have also verified (\ref{lerch-expansion}) by direct numerical evaluation, comparing it to the results of (\ref{lerch-function}).},  
\begin{multline}
		\Phi(e^{-\xi\pi/a},1,\upsilon)e^{-\xi\pi\upsilon/a}+\log(\xi\pi/a)=-\gamma-\psi(\upsilon)\\
		-\sum_{m=0}^{\infty}(-1)^{m+1}\frac{B_{m+1}(\upsilon)}{m+1}\frac{(\pi\xi/a)^{m+1}}{(m+1)!}\label{lerch-expansion}
\end{multline}
where \(\gamma\) is Euler's constant, \(\psi(a/2L)\) the digamma function (logarithmic derivative of the gamma function), and the \(B_{m+1}(a/2L)\) are Bernoulli polynomials~\cite{bateman1953}.  From (\ref{lerch-expansion}) we find, after neglecting positive powers of \(\xi\),
	\begin{multline}
		\frac{1}{\xi}\frac{d^{2}}{d\xi^{2}}\left[\Phi(e^{-\xi\pi/a},1,\upsilon)-\Phi(e^{-\xi\pi/a},1,-\upsilon)\right]\sim-\frac{2\pi\upsilon}{a\xi^{2}}\\
		-2\left(\frac{\pi\upsilon}{a}\right)^{3}\log{(\xi\pi/a)}-\frac{\pi^{2}\upsilon}{a^{2}\xi}\left[1+\upsilon(\psi(\upsilon)-\psi(-\upsilon))\right]\\
		-\left(\frac{\pi}{a}\right)^{3}\upsilon\left\{\upsilon^{2}[2(\gamma-1)+\psi(\upsilon)+\psi(-\upsilon)]+1/6\right\}\label{lerch-expansion-2}
	\end{multline}
Applying (\ref{lerch-expansion-2}) and (\ref{expansion-1}) to (\ref{energy-derivative}), gives the following lengthy expression for the rate of change of energy with respect to \(\alpha\),
	\begin{widetext}
		\begin{multline}
			\frac{1}{A}\frac{d\tilde{E}}{d\alpha}=\frac{\hbar c}{4\pi^{2}}\left[1-\cos(\pi a/L)\right]\bigg\{-\frac{6L}{\pi\xi^{4}}-\frac{L}{a\xi^{3}}-\frac{\pi}{4L\xi^{2}}-\frac{\pi^{2}}{8aL\xi}\left[1+\frac{a}{2L}\left(\psi(a/2L)-\psi(-a/2L)\right)\right]-\frac{\pi^{3}}{16L^{3}}\log(\xi\pi/a)\\
			-\frac{\pi^{3}}{8a^{2}L}\left[\left(\frac{a}{2L}\right)^{2}\left[2(\gamma-1)+\psi(a/2L)+\psi(-a/2L)\right]+1/6\right]+\frac{L\pi^{3}}{360a^{4}}\bigg\}+a\to L-a.\label{rate-of-change}
		\end{multline}
	\end{widetext}
	Evidently the divergent terms---except those proportional to \(\xi^{-2}\) and \(\xi^{-4}\)---are dependent on the position of the piston within the chamber. These divergences are not cancelled by including the contributions to the energy from the other side of the piston.  This means that in the limit \(\xi\to0\), the force on the piston is discontinuous as a function of \(\alpha\), being finite when \(\alpha=0\), and infinite as \(\alpha\) is moved away from zero.  It therefore seems that a \(\xi\) independent meaning cannot be given to the Casimir force, when in an inhomogeneous medium. The value of the force is cutoff-dependent.
%
%: Relation to previous proposals
%
	\section{Relation to previous proposals\label{end-section}}
	\par
	It was recently proposed~\cite{goto2012} that one should calculate the Casimir force in an inhomogeneous medium through forming a Laurent expansion of the energy in powers of the regularizing parameter (here \(\xi\)).  The regularized energy is then defined to be ``\emph{the term, \(c_{0}\) in} [the] \emph{Laurent expansion that is independent of} [\(\xi\)] \emph{and corresponds to discarding the principal part of the Laurent series before taking the limit as} [\(\xi\)] \emph{tends to zero}''.  This procedure works well for homogeneous systems. However, applying this procedure to (\ref{rate-of-change}) (which is proportional to the change in energy due to the inhomogeneous medium, when \(\alpha\) is small), proves problematic. For one thing,  we would still be left with the logarithmic divergence.  More seriously, however, the negative powers of \(\xi\) which depend on \(a\) are simply swept away.  Therefore the regularization procedure modifies the value of the force on the mirrors.  There is no obvious physical justification for this and we emphasize, as we did in section~\ref{section-1}, that the regularization in the case of a homogeneous medium does not modify the value of the force (the diverging terms are independent of the positions of the mirrors).
%
%: Conclusions
%
	\section{Conclusions}
	\par
	It is has been demonstrated in previous work~\cite{simpson2013} that there is a problem with the Lifshitz theory of Casimir forces: the Casimir-stress in inhomogeneous media, where the optical properties of the medium vary continuously in space, is infinite and resists regularisation. But the problem of computing Casimir forces in inhomogeneous media appears to be systemic.  Here we have shown that, even with a simple energy mode summation, the ground-state energy of a system proves similarly resistive to regularisation. We have shown this for the case of a Casimir piston, when an inhomogeneous medium is present in the cavity. Our calculation demonstrates a cutoff dependence in the Casimir force which suggests, surprisingly, that the Casimir forces in a system depend in detail on its microphysical properties.  If this is the case, it seems unlikely that a generally finite and physically meaningful result could be obtained through a simple modification to the existing regularisation procedure~\cite{leonhardt2009,goto2012,leonhardt2011}; some additional physics must be taken into account.
%
%: Acknowledgements
%
	\acknowledgements
	The authors wish to thank T. G. Philbin, U. Leonhardt and Y. Imry for useful discussions.  SARH thanks the EPSRC for financial support, and the Weizmann Institute for their hospitality.  WMRS thanks SUPA and the Weizmann Institute for their financial support.
%
%: Appendices
%
\pagebreak
\appendix
\begin{widetext}
\section{Integration of perturbed eigenfrequencies\label{appendix-A}}
\par
	Here we give the results of the following integrals,
\begin{equation}
	I^{s}_{m,\lambda}=\int_{0}^{\infty}k_{\parallel}dk_{\parallel}\omega^{s\,(1)}_{m,\lambda}e^{-\xi\omega^{s\,(0)}_{m,\lambda}/c},\label{integral}
\end{equation}
	where the \(\omega^{s\,(1)}_{m,\lambda}\) are given by (\ref{left-perturbation-1}--\ref{left-perturbation-2}), with the \(s=`R'\) expressions are obtained after the substitution \(a\to L-a\).  The eigenfrequency shifts for polarization, \(\lambda=1\) depends on \(k_{\parallel}\) only through \(\omega^{s\,(0)}_{m,1}\): i.e. in the same way as the unperturbed eigenfrequencies.  Therefore (\ref{integral}) can be performed for this polarization in the same way as section~\ref{section-1}, with the result,
	\begin{equation}
		I^{L}_{m,1}=-\frac{\alpha cL\left[1-\cos(\pi a/L)\right]}{\pi a}\left[-\frac{1}{\xi^{3}}+\frac{1}{\xi^{2}}\frac{d}{d\xi}-\frac{1}{2\xi}\frac{d^{2}}{d\xi^{2}}\right]\frac{(2mL/a)^2}{1-(2mL/a)^2}e^{-\xi m\pi/a}.\label{integral-1}
	\end{equation}
	Polarization, \(\lambda=2\), has a slightly more complicated dependence on \(k_{\parallel}\).  Yet there is nothing fundamentally different about performing the integrals, and one obtains,
	\begin{equation}
		I^{L}_{m,2}=\frac{\alpha cL[1-\cos(\pi a/L)]}{\pi a}\bigg\{\left[\frac{1}{1-(2mL/a)^2}+1\right]\left(\frac{1}{\xi^{2}}\frac{d}{d\xi}-\frac{1}{\xi^{3}}\right)+\frac{(2mL/a)^2}{1-(2mL/a)^2}\frac{1}{2\xi}\frac{d^{2}}{d\xi^{2}}\bigg\}e^{-\xi m\pi/a}\label{integral-2}.
	\end{equation}
	After substituting \(a\to L-a\) into (\ref{integral-1}) and (\ref{integral-2}), one obtains \(I^{R}_{m,1}\) and \(I^{R}_{m,2}\).
\end{widetext}
\bibliography{refs}
\end{document}